\documentclass[preprint,showpacs,preprintnumbers,amsmath,amssymb]{revtex4}

\usepackage{graphicx}% Include figure files
\usepackage{dcolumn}% Align table columns on decimal point
\usepackage{bm}% bold math

\begin{document}

%\preprint{APS/123-QED}

%\begin{document}

% Use the \preprint command to place your local institutional report
% number in the upper righthand corner of the title page in preprint mode.
% Multiple \preprint commands are allowed.
% Use the 'preprintnumbers' class option to override journal defaults
% to display numbers if necessary
%\preprint{}

%Title of paper
\title{ On the Hylleraas Coordinates }

% repeat the \author .. \affiliation  etc. as needed
% \email, \thanks, \homepage, \altaffiliation all apply to the current
% author. Explanatory text should go in the []'s, actual e-mail
% address or url should go in the {}'s for \email and \homepage.
% Please use the appropriate macro foreach each type of information

% \affiliation command applies to all authors since the last
% \affiliation command. The \affiliation command should follow the
% other information
% \affiliation can be followed by \email, \homepage, \thanks as well.
\author{Xiao-Yin Pan}
\author{ Viraht Sahni}
%\email[]{Your e-mail address}
%\homepage[]{Your web page}
%\thanks{}
%\altaffiliation{}
\affiliation{ The Graduate School of the City University of New
York, New York, New York 10016. }
\author{Lou Massa}
\affiliation{ The Graduate School of the City University of New
York, New York, New York 10016. \\
 Crest Center for
Mesoscopic Modeling and Simulation,The City University of New
York, New York 10016}
%Collaboration name if desired (requires use of superscriptaddress
%option in \documentclass). \noaffiliation is required (may also be
%used with the \author command).
%\collaboration can be followed by \email, \homepage, \thanks as well.
%\collaboration{}
%\noaffiliation

\date{\today}

\begin{abstract}
The Hylleraas coordinates $s=r_{1}+r_{2}$, $t=r_{1}-r_{2}$,
$u=|{\bf r}_{1}-{\bf r}_{2}|$ are the natural coordinates for the
determination of properties of the Helium atom, the positive ions
of its isoelectronic sequence, and the negative Hydrogen ion.  In
this paper, we derive a new expression for integrals representing
properties such as the energy, normalization and expectation of
arbitrary operators, as written in the $(s,t,u)$  coordinates. The
expression derived is valid for both \emph{finite} and
\emph{infinite} space. The integrals for the various properties
are comprised in each case of two components $A$ and $B$. The
contribution of these components to the volume of integration and
the normalization of a wave function for finite space, and in
variational calculations of the ground state energy of the Helium
atom confined in a finite volume is demonstrated by example. We
prove that when the integration space is \emph{infinite}, the
expression for the energy and other properties employed by
Hylleraas corresponds \emph{only} to that of  integral $A$. We
further prove that for the approximate variational wave functions
employed by Hylleraas and other authors, the contribution of the
term $B$ vanishes. This contribution also vanishes for the exact
wave function. It is interesting to note that the component $B$ to
the integral is not mentioned in the literature. A principle
purpose of the paper, therefore, is to point  out the existence of
this term.

\end{abstract}

% insert suggested PACS numbers in braces on next line
%\pacs{}
% insert suggested keywords - APS authors don't need to do this
%\keywords{}

%\maketitle must follow title, authors, abstract, \pacs, and \keywords
\maketitle

% body of paper here - Use proper section commands
% References should be done using the \cite, \ref, and \label commands
\section{ Introduction}
The Schr{\"o}dinger equation \cite{1} for the Helium atom, or the
positive ions of its isoelectronic sequence, or of the negative
Hydrogen ion, is
\begin{equation}
{\hat H}\Psi=E \Psi,
\end{equation}
where the Hamiltonian ${\hat H}$ is
\begin{equation}
\hat{H}=-\frac{1}{2}\nabla_{1}^{2}-\frac{1}{2}\nabla_{2}^{2}-\frac{Z}{r_{1}}
  -\frac{Z}{r_{2}}+\frac{1}{r_{12}},
\end{equation}
where ${\bf r}_{1}$, ${\bf r}_{2}$ are the coordinates of the two
electrons, $r_{12}$ is the distance between them(see Fig.1), $Z$
 the charge of the nucleus, and $\Psi$ and $E$ the eigenfunction
 and eigenenergy of the system. As noted by Hylleraas \cite{2},
 the Schr{\"o}dinger equation in this instance is not a six-dimensional
 but rather a three-dimensional eigenvalue problem. The natural
 coordinates of the system are ${ r}_{1}$, ${ r}_{2}$, and
  ${ r}_{12}$
 forming the triangle $({ r}_{1}, {r}_{2},  { r_{12}} )$.
 The other three angle variables which determine the orientation of
 the triangle in space, because of the symmetry of the system,
  are arbitrary and can be integrated over. Equivalently, the
  system is uniquely described by the coordinates ${ r}_{1}$, ${ r}_{2}$,
  and $\theta_{12}$, the angle between the vectors ${\bf r}_{1}$ and
  ${\bf r}_{2}$.\\

 As is well known, there is no closed-form analytical
 solution to the two-electron eigenvalue problem. Hence,
 the energy is most accurately obtained by variational
 minimization of the energy functional $E[\Psi]$ defined as
\begin{equation}
E[\psi]=\frac{\int \psi {\hat H} \psi d\tau }{\int \psi^{2}d\tau},
\end{equation}
with respect to the parameters in the wave function. In performing
such variational calculations, Hylleraas further introduced the
`elliptical' coordinates
\begin{equation}
s=r_{1}+r_{2}, \; t=r_{1}-r_{2}, \;  \; u=r_{12}.
\end{equation}
Then assuming approximate wave functions dependent on these
coordinates, he applied the variational principle to obtain the
energy. There has been similar work employing these $(s,t,u)$
coordinates over the past decades \cite{3,4,5,6,7,8}.\\

In this paper, we investigate integrals of the form $\int f d\tau
$ appearing in Eq. (3) in the elliptical coordinates $(s,t,u)$ for
both finite and infinite space. We have discovered another way of
writing the integrals in these coordinates that differs from the
commonly employed one in the literature. In section II we derive
the integral to be a sum of two contributions $A$ and $B$ valid
for both \emph{finite} and \emph{infinite} space. This is a
natural separation of the integration domain.  We demonstrate the
correctness of our derivation by three examples for finite space
as described in section III. These examples correspond to (a)
volume of integration, (b) normalization of a wave function, and
(c) variational calculations of the ground state energy of the
Helium atom confined to a finite volume. In section IV, we prove
for the case when the space is \emph{infinite} that the energy and
normalization integrals employed by Hylleraas and others
corresponds \emph{only } to that of integral $A$. We further prove
that for the approximate wave functions employed by these authors,
the contribution of  $B$ vanishes. As the asymptotic structure of
the exact wave function is known to also decay
exponentially\cite{9}, once again the contribution of integral $B$
for infinite space vanishes for the true wave function. We
summarize our results and conclusions
in section V. \\

\section{ INTEGRALS IN $(s,t,u)$ COORDINATES }
 \begin{figure}
 \begin{center}
 \includegraphics[bb=14 12 420 362,angle=0, scale=0.8]{Fig1.eps}
 \caption{The coordinate system indicating the positions ${\bf r}_{1}$ and ${\bf r}_{2}$
 of the two electrons with the nucleus at the origin. \label{}}
 \end{center}
 \end{figure}
Let us first focus on the volume element $d\tau$  of Eq.(3).
Employing the symmetry  of the system, $d\tau$ can be rewritten as
\begin{equation}
d\tau=d{\bf r}_{1}d{\bf
r}_{2}=J(r_{1},r_{2},\theta_{12})dr_{1}dr_{2}d\theta_{12},
\end{equation}
where $J(r_{1},r_{2},\theta_{12})$ is the Jacobian of the
coordinate transformation. By \emph{fixing} ${\bf r}_{1}$ first,
carry out the integral of ${\bf r}_{2}$ with respect to ${\bf
r}_{1}$ to arrive at (see Fig.1)
\begin{equation}
\int d\tau=\int d{\bf r}_{1} r_{2}^{2}dr_{2} \; sin\theta_{12}
d\theta_{12} d\varphi_{12}=2 \pi \int d{\bf r}_{1}r_{2}^{2}dr_{2}
\;sin\theta_{12} d\theta_{12}.
\end{equation}
Next, by integrating over ${\bf r}_{1}$, we note there is no
dependence on the Euler angles, so that Eq.(6) reduces to
\begin{equation}
 \int d\tau =8 \pi^{2} \int r_{1}^{2} r_{2}^{2} dr_{1}dr_{2} \; sin\theta_{12}
d\theta_{12}.
\end{equation}\\

We now introduce the new variable $u$, the distance between ${\bf
r}_{1}$ and ${\bf r}_{2}$:
 \begin{equation}
 u^{2}=r_{12}^{2}=r_{1}^{2}+r_{2}^{2}-2 r_{1}r_{2} \; cos\theta_{12}.
 \end{equation}
During the previous integration steps, \emph{if  $r_{1}$ and
$r_{2}$ are fixed first}, then
\begin{equation}
u\;du=r_{1}r_{2} \; sin\theta_{12}d\theta_{12},
\end{equation}
so that (for the case of \emph{infinite} volume)
\begin{equation}
 \int d\tau =8 \pi^{2} \int_{0}^{\infty} \int_{0}^{\infty}r_{1}^{2} r_{2}^{2} dr_{1}dr_{2} \;\int_{0}^{\pi} sin\theta_{12}
d\theta_{12}=8 \pi^{2} \int_{0}^{\infty} \int_{0}^{\infty} r_{1}
r_{2} dr_{1}dr_{2} \;\int_{|r_{1}-r_{2}|}^{r_{1}+r_{2}} u du.
\end{equation}
Eq.(10) appears in the work of Hylleraas\cite{2}.\\

 \begin{figure}
 \begin{center}
 \includegraphics[bb=0 0 308 261, angle=0, scale=0.8]{Fig2.eps}
 \caption{The domain $S$ of integration of the coordinates $r_{1}$ and $r_{2}$.\label{}}
 \end{center}
 \end{figure}

Let us next  confine the two electrons in some\emph{ finite}
volume of space such that  $0\leq r_{1}\leq R$ and  $0\leq
r_{2}\leq R$. The reason we choose the same $R$ is because of the
symmetry between the two electrons.  The limit $R\longrightarrow
\infty$ then leads to the infinite space integral. For the
elliptical coordinates $s$ and $t$ of Eq.(4), it is easy to show
that
\begin{equation}
dr_{1} dr_{2}=\frac{1}{2}ds dt ,\; r_{1}
r_{2}=\frac{1}{4}(s^2-t^2).
\end{equation}
Therefore, for any integrand  $f$,  which  is a function of
$(s,t,u)$, we have from Eqs.(10) and (11) that the integral in
terms of the new coordinates $(s,t,u)$ is
\begin{equation}
\int f  d\tau =8\pi^{2} {\int_{0}^{R}\int_{0}^{R}} r_{1}
r_{2}dr_{1}dr_{2} \int_{|r_{1}-r_{2}|}^{r_{1}+r{2}} u f(s,t,u)
du\\
=8\pi^{2} {\int\int}_{S'} \frac{1}{2}ds dt \int_{|t|}^{s} f(s,t,u)
\frac{s^{2}-t^{2}}{4} u du.
\end{equation}
Here $S'$ denotes the integration domain in the new coordinates
$(s,t)$. This domain  has changed from $S:[0\leq r_{1}\leq R,
0\leq r_{2}\leq R]$ as shown in Fig.2 to $S':[0\leq s\leq 2 R;
-R\leq t\leq R]$.\\

\begin{figure}
 \begin{center}
 \includegraphics[bb=0 0 317 284, angle=0, scale=0.8]{Fig3.eps}
 \caption{The domain $S'$ of integration of the Hylleraas  coordinates $s$ and $t$.\label{}}
 \end{center}
 \end{figure}

The function $f(s,t,u)$  could be $ \psi {\hat H} \psi $ or
$\psi^{2}$. As stated  by Bethe\cite{10}: `` The exact symmetry
requirement [of the wave function] then takes the simple form that
$\psi$ be an even function of $t$ for parahelium, an odd function
of $t$ for ortho-helium. Since the Hamiltonian is an even function
of $t$ and since the integrals in Eq.(3) contain two factors, the
contribution to the integral from $-t$ is identical with that from
$+t$. We therefore restrict ourselves to positive values of t in
the integrals and multiply the volume element by a factor of
$2$''. With that in mind, Eq. (12) can be rewritten as
\begin{eqnarray}
 \int f  d\tau &=&2\pi^{2} {\int\int}_{S'}ds dt \int_{|t|}^{s} f(s,t,u)
(s^{2}-t^{2}) u du \nonumber \\
&=&2\pi^{2} \int_{0}^{R} ds \int_{0}^{s} dt\int_{t}^{s}
f(s,t,u)(s^{2}-t^{2}) u du \nonumber \\
&+&2\pi^{2} \int_{R}^{2R} ds \int_{0}^{2R-s}dt \int_{t}^{s}
f(s,t,u)(s^{2}-t^{2}) u du.
\end{eqnarray}\\

Now we can see from equation (13) that the integral has two
components. For convenience, let us denote them as $A$ and $B$
where
\begin{equation}
A=2\pi^{2} \int_{0}^{R} ds \int_{0}^{s} dt \int_{t}^{s}
f(s,t,u)(s^{2}-t^{2}) u du,
\end{equation}
and
\begin{equation}
B=2\pi^{2} \int_{R}^{2R}ds \int_{0}^{2R-s} dt\int_{t}^{s}
f(s,t,u)(s^{2}-t^{2}) u du.
\end{equation}\\

We next give examples to demonstrate the correctness of this
separation of the volume of integration.\\

\section{ Examples  }

(a) \emph{Volume of integration}

   We first consider the volume of integration by choosing
   $f(s,t,u)=1$. The result should be the square of the volume of  a sphere with
radius $R$. Indeed, one easier way to determine this is by
employing Eq.(7)
\begin{equation}
\int d\tau=8 \pi^{2}\cdot 2 (\frac{R^{3}}{3})^{2}=(\frac{4 \pi
R^{3}}{3})^{2}.
\end{equation}
On the other hand, we can carry out the integral via its
components  derived  in Eq.(13). On using the following integrals
\begin{equation}
\int_{t}^{s}(s^{2}-t^{2}) u du=\frac{(s^{2}-t^{2})^{2}}{2},
\end{equation}
\begin{equation}
\int_{0}^{R} ds\int_{0}^{s}du \frac{(s^{2}-t^{2})^{2}}{2}=\frac{2
R^{6}}{45},
\end{equation}
\begin{equation}
\int_{R}^{2R} ds \int_{0}^{2R-s}du
\frac{(s^{2}-t^{2})^{2}}{2}=\frac{38R^{6}}{45},
\end{equation}
 we also have
\begin{equation}
 \int d \tau=A+B= 2\pi^{2} (\frac{2
R^{6}}{45}+\frac{38R^{6}}{45})=(\frac{4 \pi R^{3}}{3})^{2}.
\end{equation}
The agreement of the two ways to obtain the volume demonstrates
that the new way of expressing the integrals within the $(s,t,u)$
coordinates as defined by  Eq.(13) is correct.\\

(b) \emph{Normalization}

Let us next consider the normalization of the trial wave function
\begin{equation}
 \psi=C e^{- \alpha s},
\end{equation}
where $\alpha$ is a variational parameter, $C$ is the
normalization constant. Then in spherical polar coordinates for a
finite volume of radius $R$, the normalization integral is
\begin{eqnarray}
1=\int \psi^{2} d\tau &= &C^{2} (4 \pi \int_{0}^{R} e^{-2
\alpha r}r^{2} dr )^{2}  \nonumber \\
&=& \frac{C^{2}\pi^{2}}{\alpha^{6}} [ 1-e^{-2 \alpha R}(1+2 \alpha
R+2 \alpha^{2} R^{2} )]^{2}.
\end{eqnarray}
For $R\rightarrow \infty$, $C=\alpha^{3}/\pi$, the well known
result.\\

In the (s,t,u) coordinates, the contribution from $A$ of Eq.(14)
is
\begin{eqnarray}
A&=&\int_{A} \psi^{2} d\tau =2 \pi^{2} C^{2} \int_{0}^{R} ds e^{-2
\alpha s}\int_{0}^{s} dt \int_{t}^{s} du u (s^{2}-t^{2}) \\
&=& \frac{C^{2}\pi^{2}}{\alpha^{6}} [1-\frac{e^{-2 \alpha
R}}{15}(15+30 \alpha R+30 \alpha^{2} R^{2}+20 \alpha^{3} R^{3}+10
\alpha^{4} R^{4}+4 \alpha^{5} R^{5})].
\end{eqnarray}
The contribution to normalization from B of Eq.(15) is
\begin{eqnarray}
B&=&\int_{B} \psi^{2} d\tau =2 \pi^{2} C^{2} \int_{R}^{2R} ds
e^{-2
\alpha s}\int_{0}^{2R-s} dt \int_{t}^{s} du u (s^{2}-t^{2}) \\
&=& \frac{C^{2}\pi^{2}}{\alpha^{6}} [e^{-4 \alpha R}(1+2 \alpha
R+2 \alpha^{2} R^{2})^{2} \nonumber \\
& & -\frac{e^{-2 \alpha R}}{15}(15+30 \alpha R+30 \alpha^{2}
R^{2}-20 \alpha^{3} R^{3}-10 \alpha^{4} R^{4}-4 \alpha^{5}
R^{5})].
\end{eqnarray}
It is easily verified that the sum of $A$ and $B$ of Eq.(24) and
Eq.(26) are equivalent to that of Eq.(22). This again demonstrates
the correctness of the two integrals derived.\\

(c) \emph{Variational Calculations}

In the final example,  we consider the Helium atom confined within
shells of volume of integration corresponding to $R=2.0, 3.0,
4.0$, and $5.0$ (a.u.). We determine variationally the ground
state energy of the atom thus confined. The approximate wave
function  we employ is that of Eq.(21), which then does not
satisfy the boundary condition of vanishing at the surface.
 The total energy of Eq.(3) is determined in both
spherical polar coordinates as well as via Eq.(14) and Eq.(15) of
contributions $A$ and $B$. A point to note is that the commonly
employed expression for the integrand of the kinetic energy in
(s,t,u) coordinates is valid only for infinite space and assumes
that the wave function vanishes there. However, since the volume
of integration is finite, there is a contribution to the kinetic
energy  from the surface of the volume as the wave function does
not vanish on the surface (see the Appendix for details). The
analytical expression for the kinetic energy contributions from
integrals $A$ and $B$ and the surface contribution as well as
those for the total energy are given in the Appendix. The various
numerical contributions of the kinetic energy $T$, external
$E_{ext}$ and electron-interaction $E_{ee}$ potential energies are
listed in Table I. The well known\cite{10} variational result  for
\emph{infinite} space quoted to six significant figures  is
$E=-2.84765$ (a.u.) for $\alpha=27/16$. For each value of $R$, the
sum of the contributions from integrals $A$ and $B$ for the
\emph{separate} kinetic plus surface $S$ contribution, external,
and electron-interaction energy components
is the same as those obtained in spherical polar coordinates.\\

Note there is a contribution to each component of the energy from
both integral $A$ and $B$ for each value of $R$. Due to the
exponential decay of the wave function, the contribution from $B$
diminishes but always is finite even at $R=5.0$ (a.u.). Observe
that at $R=5.0$ (a.u.), the total energy is the same as the well
known result to six significant figures. However, the contribution
of $B$ becomes negligible in the $R\rightarrow \infty $ limit.
Thus, it is only in this limit that the integral corresponding to
$B$ can be ignored. The variational results once again prove the
correctness of the derivations for the two
integrals $A$ and $B$.\\

\begin{table}
\caption{\label{tab:table I} The variationally determined ground
state energy of the Helium atom and of the corresponding kinetic
energy $T$, external $E_{ext}$ and electron-interaction $E_{ee}$
potential energy components as a function of the radius $R$ of the
volume of integration in (a.u.).
 The contribution  from the integrals $A$ and $B$
 for each energy component as well as the surface contribution to the kinetic energy are quoted as are
 the energy minimized values of the variational parameter $\alpha$.}

\renewcommand{\arraystretch}{1.2}
%\begin{ruledtabular}
\begin{tabular}{||c |c |c |c| c|c| c|c| c|c||}
\hline \hline
 $R$ & $\alpha$ & \multicolumn{3}{c|}{$T$ } & \multicolumn{2}{c|}{{$E_{ext}$}}& \multicolumn{2}{c|}{ $E_{ee}$ }& $E$\\
 &   & \multicolumn{3}{c|}{}  & \multicolumn{2}{c|}{}& \multicolumn{2}{c|}{} & \\ \cline{3-9}
&  &  A &  B & S &  A &  B &  A &  B &  \\
\hline

 $2.0$ & $1.652$ & $1.91327$ & $0.81583$ & $0.16747$ &
$-5.64802$& $-1.16273$ & $-0.88250$ & $0.20120$ & $-2.83048$
\\\hline

 $3.0$ & $1.687$ & $2.68121$ & $0.16476$ & $0.01175$ &
$-6.59864$& $-0.16328$ & $1.03104$ & $0.02713$ & $-2.84603$
\\\hline

 $4.0$ & $1.688$ & $2.82820$ & $0.02114$ & $0.00071$ &
$-6.73641$& $-0.01643$ & $1.05256$ & $0.00267$ & $-2.84756$
\\\hline

 $5.0$ & $1.688$ & $2.84729$ & $0.00205$ & $0.00004$ &
$-6.75073$& $-0.00131$ & $1.05480$ & $0.00021$ & $-2.84765$
\\\hline

\hline \hline
\end{tabular}
%\end{ruledtabular}
\end{table}

\section{ COMPARISON WITH THE HYLLERAAS INTEGRAL}
In this section, we compare our new way of writing the integral
with that of Hylleraas.  As noted previously, Eq. (10) appears in
the work of Hylleraas. Following this equation, and without any
further detail, he then writes the integral for the energy and
normalization for when the space is \emph{infinite} as
\begin{equation}
\int f d\tau=2 \pi^{2} \;\int_{0}^{\infty}ds \int_{0}^{s} du
\int_{0}^{u}dt \; u \;(s^{2}-t^{2})\;f(s,t,u).
\end{equation}
Observe that in this integral, the integration over the variable
$t$ is performed before that of the variable $u$.  (We note that
in all subsequent literature employing these elliptical
coordinates, it is the Hylleraas expression that is employed.)
This order of integration is surprising because the variable $u$
depends on the variables $s$ and $t$.  Hence, it is natural to
perform the integral over $u$ prior to that of $t$ and $s$ as in
our expression Eq. (13). The derivation of Eq. (13) is a
consequence of our attempt to understand how Hylleraas arrived at
his expression of Eq. (27). For a comparison of Eqs. (13) and
(27), we take  the $R\longrightarrow \infty$ limit of Eq.(13). The
two results ought to be equivalent. However, we find that it is
only $lim_{R\longrightarrow \infty } A$ (see Eq.(14)) that is
equivalent to Eq.(27) of Hylleraas.
 For $R\longrightarrow \infty$
\begin{equation}
A|_{R\longrightarrow \infty}=2 \pi^{2} \;\int_{0}^{\infty}ds
\int_{0}^{s} dt \int_{t}^{s}du \; u \;(s^{2}-t^{2})\;f(s,t,u).
\end{equation}
 As shown by Fig.4, in Eq.(27), for each $u$ with $0\leq u\leq s$,
 we have
 $0\leq t\leq u$ , the area swept is the upper shaded
 triangle. On the other hand, in Eq.(28), for each $t$ with
 $0\leq t\leq s$, we have $t\leq u\leq s$ , so that the area
  is the  same as in
 Eq.(27). Therefore, for any integrand, the integrals of Eq.(27)
  and (28) are the
 same. \emph{Thus, the Hylleraas integral is the same as
 that of $A$ for $R\longrightarrow \infty$}. The difference in the order of integration
 between the expression of Hylleraas and of our derivation is of critical significance for work
 on the Helium atom to be published elsewhere.\\

 \begin{figure}
 \begin{center}
 \includegraphics[bb=0 0 323 272, angle=0, scale=0.8]{Fig4.eps}
 \caption{The domain  of integration of the Hylleraas  coordinates $t$ and $u$.\label{}}
 \end{center}
 \end{figure}

 Let us now turn to Eq.(15) and examine the integral $B$ in the same limit.
 As we have shown earlier, in finite space, the contribution of $B$
 plays an important role. When $R\longrightarrow \infty$, we have
\begin{eqnarray}
B|_{R\longrightarrow \infty}&=&2 \pi^{2} lim_{R\longrightarrow
\infty} \int_{R}^{2R}ds \int_{0}^{2R-s}dt \int_{t}^{s}
f(s,t,u)(s^{2}-t^{2})udu \\
&=& lim_{R\longrightarrow \infty} \; 2\pi^{2} \int_{R}^{2R}
h(R,s)ds
\end{eqnarray}
\emph{It is only when the function $h(R,s)$ decays in a manner
such that $lim_{R\longrightarrow \infty}B$ vanishes that the
Hylleraas expression corresponds to the exact value of the
integral $\int f d\tau$}. Thus, for example, if one assumes as in
all prior literature, that $h(R,s)$  decays exponentially, and is
of the form
\begin{equation}
h(R,s)=g(R) e^{- \alpha s}  \sum_{l \geq 0} s^{l},
\end{equation}
then since for any non-negative polynomial of $s$,
\begin{equation}
lim_{R\longrightarrow \infty} \;  \int_{R}^{2R}ds e^{- \alpha s}
s^{l}=0,
\end{equation}
we have $B|_{R\longrightarrow \infty}=0$. However, if the
dependence on the coordinate $s$ contains  terms of the form
$s^{p}$  with $p\geq -1$ or is multiplied by some power of $lns$,
the contribution of $B$ could be nonzero. For example, the
integral $\int_{R}^{2R} h(s) ds $ for $h(s)=1/s$ is $ln2 $
irrespective of how large $R$ is. As noted above, for the form of
approximate wave functions employed in the literature, the
contribution of $B$ always vanishes in the limit as $R\rightarrow
\infty$. It also vanishes for the exact wave function whose
asymptotic decay is known \cite{9} to be $r^{\beta}e^{-\alpha r}$,
where $(1+\beta)=(Z-N+1)/\alpha$, and $\alpha=\sqrt{2I}$, with $I$
being the ionization potential.\\

It is interesting that the contribution of the integral $B$ to the
integral $\int f d\tau $ is not mentioned at all in the
literature.  One would imagine that in spite of the fact that the
contribution of $B$ vanishes for the choice of a particular
approximate wave function, mention of this term pointing out its
lack of contribution would appear somewhere.  But this is not the
case. Hence, one principal purpose of this
paper is to note the existence of this term.\\

 \section{ CONCLUSION}
  In this paper we have derived a new expression for integrals representing the
energy, normalization, or other expectation values when written in
the Hylleraas  $(s,t,u)$  coordinates, i.e. $\int f(s,t,u)d\tau$.
The expression derived for the intergral is valid for both finite
and infinite
 space. The integrals for the various properties are comprised in
 each case of two components $A$ and $B$. The contribution of
 these components to the volume of integration, the normalization
 of a wave function, and in variational calculations of the
 energy, for finite space is demonstrated by example.
  We prove that when the space of integration is infinite, the
  expression used by Hylleraas and others corresponds \emph{only} to that of
integral $A$. In the literature,  the form of the approximate wave
functions are such that the integrands for the energy and
normalization  are usually of the form $e^{- \alpha s}\sum_{l,m,n}
h_{l}(s) t^{m}u^{n}$; $h_{l}(s)=s^{l}$ or $(ln s)^{l}$
\cite{7,11}, etc. We show that for such wave functions the
contribution of $B$ vanishes. The contribution of integral  $B$
also vanishes for the exact wave function. Thus,  in calculations
of the integral $\int f(s,t,u)d\tau $ when the space considered is
infinite, one may ignore the contribution of the component $B$ by
choosing an appropriate decay of the approximate wave function.
 Nevertheless, we reiterate that from a rigorous mathematical perspective,
  the integral $\int f(s,t,u)d\tau$ is composed of two components $A$ and $B$.\\

\appendix*
\section{}

 For the wave function of Eq.(21), the contribution to the total
 energy from integral $A$ is
 \begin{equation}
 E_{A}=2 \pi^{2} \int_{0}^{R}ds \int_{0}^{s} dt \int_{t}^{s} du [u (s^{2}-t^{2}) (\frac{\partial \psi}{\partial s})^{2}-4 Z s
 u C^{2}e^{-2 \alpha s}+(s^{2}-t^{2})C^{2} e^{-2 \alpha s})],
\end{equation}
 where the first, second, and third terms correspond
respectively to the kinetic, external, and electron-interaction
energies. (The surface contribution to the kinetic energy is
discussed below). Thus,
\begin{eqnarray}
E_{A}&=&\frac{\pi^{2}C^{2}}{8 \alpha^{5}}
[-(27-8\alpha)-\frac{1}{15}\{e^{-2 \alpha R}[-405-30(-4+27
R)\alpha -30 R(-8+27 R)\alpha^{2}\nonumber \\
& &-60 R^{2} (-4+9 R)\alpha^{3} +10 (16-27R)R^{3}\alpha^{4}+80
R^{4} \alpha^{5}+32R^{5}\alpha^{6}] \}].
\end{eqnarray}
The contribution from  integral  $B$ is
 \begin{eqnarray}
 E_{B}&=&2 \pi^{2} \int_{R}^{2R}ds \int_{0}^{2R-s} dt \int_{t}^{s} du [u (s^{2}-t^{2}) (\frac{\partial \psi}{\partial s})^{2}-4 Z s
 u C^{2}e^{-2 \alpha s} \nonumber \\
 & &+(s^{2}-t^{2})C^{2} e^{-2 \alpha s})]  \\
 &=&\frac{\pi^{2}C^{2}}{8 \alpha^{5}}
e^{-4 \alpha R}[-21-4(-2+21 R)\alpha +8 (4-17 R)R\alpha^{2}-32
R^{2}(-2+3R)\alpha^{3}\nonumber \\
&&+64 R^{3} \alpha^{4}+32 R^{4}\alpha^{5}]  \nonumber \\
& +&  \frac{\pi^{2}C^{2}}{120 \alpha^{5}} e^{-2 \alpha R} [315+30
(-4+21R)\alpha+ 30R(-8+5 R)\alpha^{2} -60
R^{2}(4+9R)\alpha^{3}\nonumber \\
& &+10(16-27R)R^{3}\alpha^{4}+80 R^{4}
\alpha^{5}+32R^{5}\alpha^{6}].
\end{eqnarray}

The kinetic energy is
\begin{equation}
T=-\int \psi \nabla_{1}^{2} \psi d\tau =\int (\nabla_{1}\psi
)\cdot (\nabla_{1}\psi ) d \tau- \int \nabla \cdot (\psi
\nabla_{1} \psi) d \tau.
\end{equation}
The first term of Eq.(A.5) corresponds to the first terms  in
Eqs.(A.1) and (A.3) for the total energy for integrals $A$ and
$B$, respectively. This contribution is
\begin{equation}
T_{A}+T_{B}=\alpha^{2}.
\end{equation}

The surface contribution to the kinetic energy is
\begin{eqnarray}
T_{s}&=&-\int \nabla_{1} \cdot (\psi \nabla_{1} \psi)d \tau =-\int
\nabla_{1} \cdot (e^{-\alpha r_{1}} \nabla_{1} e^{-\alpha r_{1}})d {\bf r}_{1} \nonumber \\
&=& -\int (e^{-\alpha r_{1}} \nabla_{1} e^{-\alpha r_{1}})\cdot
d {\bf S}_{1}  \\
 &=&4\pi C R^{2} \alpha e^{-2 \alpha R}=\frac{4\alpha^{4} R^{2}e^{-2 \alpha R}}{ (1-e^{-2 \alpha R}(1+2R\alpha+2R^{2}\alpha^{2}))} .
\end{eqnarray}
In spherical polar coordinates, the kinetic energy $T$ in a volume
of radius $R$ is
\begin{equation}
T=-\int \psi \nabla_{1}^{2} \psi d\tau =16 \pi^{2}C^{2}
\int_{0}^{R}\int_{0}^{R} e^{-2 \alpha (r_{1}+r_{2})}
(\alpha^{2}-\frac{2\alpha}{r_{1}}) r_{1}^{2} r_{2}^{2} dr_{1}
dr_{2},
\end{equation}
so that
\begin{equation}
T=\frac{\alpha^{2}[1+e^{-2\alpha R }(-1-2 \alpha R+2 \alpha^{2}
R^{2})] }{[1-e^{-2\alpha R }(1+2 \alpha R+2 \alpha^{2} R^{2})]}.
\end{equation}
The sum of Eq.s(A.6) and (A.8) equals Eq.(A.10).\\
%\section{}

% If you have acknowledgments, this puts in the proper section head.
\begin{acknowledgments}
This work was supported in part by the Research Foundation of
 CUNY. L. M. was supported in part by NSF through CREST, and by
 a ``Research Centers in Minority Institutions'' award, RR-03037,
 from the National Center for Research Resources, National
 Institutes of Health. We thank Professors J. B. Krieger and
 Hong-Yi Fan for their comments on the manuscript.\\
\end{acknowledgments}

\end{document}